\documentclass[12pt]{article}
\UseRawInputEncoding
\usepackage{epsfig,amsmath}

\oddsidemargin
\evensidemargin
\usepackage{graphics}

\begin{document}
\pagestyle{myheadings}
\gdef\@rtitle{}
\newcommand{\runningtitle}[1]{\gdef\@rtitle{#1}}
\gdef\@rauthor{}
\newcommand{\runningauthor}[1]{\gdef\@rauthor{#1}}
%\AtBeginDocument{
	%\markboth{\@rauthor, \@rtitle\hfil}{\hfil\@rauthor, \@rtitle}%}
\markright{\@rauthor}%
%}

%\ninept
\title{Improving D-Optimality in Nonlinear Situations}

\author{Hana Sulieman \\
Department of Mathematics and Statistics \\
American University of Sharjah, P.O.Box 26666, Sharjah, U.A.E.\\
hsulieman@aus.edu}
\date{}
\maketitle

\begin{abstract}
 Experimental designs based on the classical D-optimal criterion  minimize the volume of the linear-approximation inference regions for the parameters using local sensitivity coefficients. For nonlinear models, these designs can be unreliable because the linearized inference regions do not always provide a true indication of the exact parameter inference regions. In this article, we apply the  profile-based sensitivity coefficients developed by Sulieman et.al. \cite{sulieman1} in designing D-optimal experiments for parameter estimation in some selected nonlinear models.  Profile-based sensitivity coefficients are defined by the total derivative of the model function with respect to the   parameters. They have been shown to account for both parameter co-dependencies and model nonlinearity up to second order-derivative. This work represents a first attempt to construct experiments using profile-based sensitivity coefficients. Two common nonlinear models are used to illustrate the computational aspects of the profile-based designs and simulation  studies are conducted to demonstrate the efficiency of the constructed experiments.
\end{abstract}

\section{INTRODUCTION}
%\markright{\small D-Optimal Designs Using Profile-based Sensitivity in Regression Models}
\indent \indent Design of experiments has been a very active research area in many scientific fields for the past two decades.  For linear models, the theory of optimal designs is well established in the literature and the properties of these designs are fairly understood and used in various applications. On the other hand, for nonlinear models,  the theory of design optimality is  still emerging in the  literature. The major difficulty when the underlying model is nonlinear is that the optimal designs depend on the true value of the parameters. Hence, poor estimates of the unknown parameter values generate poor designs.

Several design of experiment techniques have been developed  and applied successfully to wide range of model systems (Franceschini and Macchietto \cite{FM08}; Berger and Wong \cite{BW09}). The objectives of these techniques typically focus on model precision or/and model discrimination. D-optimality is one of the  most popular design criteria used. The criterion minimizes the volume of the linear-approximation inference regions for the parameters. The measure of information content used in D-optimal designs involves local sensitivity coefficients defined by the first-order derivative matrix of the model function with respect to the parameters. Hence, the resulting designs are termed locally optimal. Locally optimal designs can be unreliable for highly nonlinear model functions. The works by Hamilton and Watts \cite{HW85} and Vila and Gauchi \cite{VG10} represent examples of successful attempts to take into account the model nonlinearity in the design formulation.

Sulieman et.al. \cite{sulieman1, sulieman2} proposed profile-based nonlinear sensitivity measure which simultaneously accounts for model nonlinearity and parameter estimates correlations.  Applications of the measure to different models by Sulieman et.al.\cite{sulieman3} have shown that the measure gives more reliable reflection of the sensitivity behavior of the model to the parameters than that given by the local sensitivity measures. The profile-based sensitivity measure is defined by the total derivative of the model function with respect to the parameter of interest. Hence and like any derivative measure, it is inherently local, it provides however, a broader picture of the model sensitivity in the presence of parameter co-dependencies and model nonlinearity.

The primary goal of this article is to employ the profile-based sensitivity information in the construction of D-optimal designs. The resulting designs are compared with the classical local D-optimal designs in which the conventional local sensitivity coefficients are used.
In section 2 we give a brief review of profile-based sensitivity measure and discuss its characteristics.
 In Section 3 we construct the profile-based D-optimal design and discuss its relations to the classical D-optimal design.
 Illustrative model cases are  presented
  in Section 4, and conclusions are summarized in
Section 5.
\section{A brief Overview of Profile-based Sensitivity}
Let us consider the  general mathematical form of a single response
nonlinear regression model
\begin{equation}
{\bf y}={\bf f}({\bf X},\Theta) +\mbox{\boldmath $ \epsilon $}
\label{regr}
\end{equation}
where  ${\bf y}$ is an $n$-element  vector of observed values of
the response variable for particular values of the
 $m$-regressor
variables ${\bf X}=  \{{\bf x}_{1}, {\bf x}_{2}, \ldots, {\bf
x}_{m} \}$, each ${\bf x}_{i}$ is $n$-element vector of experimental settings. $\Theta$ is  a $k$-element   vector of unknown
parameters,  ${\bf f}$ is an $n$-element  vector of  predicted
values  of the response variable for given
 ${\bf X}$ and  $\Theta$, ${\bf f}({\bf X},\Theta)=\{f({\bf x}_{1},\Theta), f({\bf
 x}_{2},\Theta),\ldots
f({\bf x}_{m},\Theta)\}$,  and $\mbox{\boldmath $ \epsilon $}$ is
an $n$-element  vector of  independent
 random errors with a specified joint distribution.
In most cases, including the case here, $\mbox{\boldmath $\epsilon
$}$ is assumed to have a spherical normal distribution, with
$E(\mbox{\boldmath $\epsilon $})= {\bf 0}$ and
$var(\mbox{\boldmath $\epsilon $})= E(\mbox{\boldmath $\epsilon
$}\mbox{\boldmath $\epsilon $}^{\prime})=\sigma^{2}{\bf I}$.

To emphasizes the dependence of the predicted response values on the parameters $\Theta$, the above model is expressed as:
\begin{equation}
{\bf y} = \mbox{\boldmath $ \eta$}(\Theta) +\mbox{\boldmath $ \epsilon $}  \label{regr1}
\end{equation}
where the $j^{th}$  element of the  $n$-dimensional
vector $\mbox{\boldmath $ \eta$}(\Theta)= (\eta_{1}(\Theta), \eta_{2}(\Theta) \\,\ldots, \eta_{n}(\Theta))^{\prime}$  is given by
\begin{equation}
\eta_{j}(\Theta)=f({\bf x}_{j},\Theta)  \ \ \ \ \ \ j=1,2,\ldots, n.
\end{equation}
Conventionally, sensitivity of model predictions to variation in parameter values is measured by the first-order partial derivative of predicted response function, $\mbox{\boldmath $ \eta$}(\Theta)$, with respect to the parameters. Sulieman {\em et al}.\cite{sulieman1} proposed  partitioning  the $p$-element parameter vector $\Theta$  into $\Theta =(\theta_{i},\Theta_{-i})$ and $\mbox{\boldmath $ \eta$}(\Theta)$ into $\mbox{\boldmath $ \eta$}(\theta_{i},\Theta_{-i})$ , where $\theta_{i}$ is the parameter of interest for which sensitivity  is measured. $\theta_{i}$ is varied across a specified range of uncertainty  so that for each fixed value of $\theta_{i}$, the conditional estimates of  the remaining  parameters
$\Theta_{-i}$ are obtained.   Accordingly a sensitivity measure  is  defined by    the total derivative of the predicted response with respect to $\theta_{i}$. It is given by:
\begin{equation}
 p_{i}({\bf x}_{0})  =\frac{D  \eta_{0}(\theta_{i}, \Theta_{-i}(\theta_{i}))}{D \theta_{i}}
 \label{PSC1}
\end{equation}
where $\eta_{0}=f({\bf x}_{0},\Theta)$ is the predicted response
at a selected point ${\bf x}_{0}$ and the notation $\displaystyle \frac{D}{D \theta_{i}}$ means total derivative with respect to $\theta_{i}$.
Based on the least squares estimation criterion, equation (\ref{PSC1}) can be expressed as:
\begin{equation}
p_{i}({\bf x}_{0})  =\frac{\partial  \eta_{0}}{\partial \theta_{i}}
-\frac{\partial  \eta_{0}}{\partial \Theta_{-i}}(\frac{\partial^{2} S}{\partial \Theta_{-i}\partial \Theta_{-i}^{\prime}})^{-1}
\frac{\partial^{2} S}{\partial \theta_{i} \partial \Theta_{-i}}\Bigr|_{\tilde{\Theta}_{-i}}
\label{eq:scnew}
\end{equation}
where the function $S$ is the sum of squares function given by $S(\Theta)= \sum_{j=1}^{n} (y_{j}-\eta_{j}( \Theta))^{2}$ and $\tilde{\Theta}_{-i}$  is the conditional least squares estimate of $\Theta_{-i}$ given $\theta_{i}$. \\
The first term in equation (\ref{eq:scnew}), $\displaystyle \frac{\partial  \eta_{0}}{\partial \theta_{i}}$ gives the conventional  sensitivity coefficient of the predicted response with respect to $\theta_{i}$. The second term is a correction term that involves the marginal effects of $\Theta_{-i}$ on $\eta_{0}$ weighted by the correlations among  the elements of $\tilde{\Theta}_{-i}$,
and  correlations between $\hat{\theta}_{i}$ and $\tilde{\Theta}_{-i}$. These parameter correlations are
based on the observed
value of the Hessian matrix, $ \displaystyle H(\Theta)=\frac{\partial^{2} S(\Theta)}{\partial \Theta \partial \Theta^{\prime}}$. $p_{i}({\bf x}_{0})$ is called {\it Profile-based Sensitivity Coefficient} after the profiling algorithm used to assess the extent of nonlinearity in the model and construct likelihood regions for parameter estimates (Bates and Watts\cite{BW88}). While $p_{i}({\bf x}_{0})$ is a local derivative-based measure, the incorporation of the correlation structure, based on the Hessian terms above,  makes it a more reliable  measure of sensitivity as it accounts for
simultaneous changes in the parameter values and nonlinearity of the model.\\
$p_{i}({\bf x}_{0})$ can be expressed in terms of the first and second order derivative information of the model function
$\mbox{\boldmath $ \eta$}(\Theta)$ as:
\begin{equation}
 p_{i}({\bf x}_{0})={v}_{0_{i}} - {{\bf v}}^{\prime}_{0_{-i}}({V}_{-i}^{\prime}{V}_{-i} -[{e}^{\prime}][{V}_{-i-i}])^{-1}
 ({V}_{-i}^{\prime}{{\bf v}}_{i}-{V_{-ii}}^{\prime}{e})
 \label{eq:psc}
\end{equation}
where ${v}_{0_{i}}$ is the $i^{th}$ component of the $k$-element first order derivative vector ${\bf v}_{0}=\displaystyle \frac{{\partial  \eta_{0}}(\Theta)}{\partial \Theta}$
evaluated at ${\bf x}_{0}$;  $V_{-i}$ is an  $n \times (k-1)$ matrix consisting of first derivative vectors of $\mbox{\boldmath $\eta$}(\Theta)$ with respect to $\Theta_{-i}$; ${{\bf v}}_{0_{-i}}$ is a $(k-1)$ dimensional vector consisting of
the elements in the row of ${V}_{-i}$ that corresponds to ${\bf x}_{0}$;
$V_{-i-i}$ is the $n \times (k-1) \times (k-1)$ array of the second order derivatives of $\mbox{\boldmath $\eta$}(\Theta)$ with respect to $\Theta_{-i}$;
$V_{-ii}$ is the $n  \times (k-1)$ matrix
of the second derivatives of $\mbox{\boldmath $\eta$}(\Theta)$ with respect to
$\Theta_{-i}$ and $\theta_{i}$, and $e$ is the $n$-element  residuals vector. In the $k-1$ parameter subspace, the entries of $[{e}^{\prime}][{V}_{-i-i}]$ give the projections of the second-order derivative vectors on the residual vector. Similarly the matrix ${V_{-ii}}^{\prime}{e}$ gives the model function curvature in $\theta_{i}$ direction  projected on the residual vector. Since the residual vector is orthogonal to the tangent plane spanned by the $k-1$ local sensitivity vectors, $[{e}^{\prime}][{V}_{-i-i}]$ is a function of only the intrinsic curvature portion of the second-order derivative array. For significant intrinsic model nonlinearity these projections play major role in the extent to which profile-based nonlinear sensitivities differ from the local linear sensitivities.
The quantities in  equation (\ref{eq:psc}) are evaluated at $(\hat{\theta}_{i}, \tilde{\Theta}_{-i}(\hat{\theta}_{i}))$.

In vector notations, equation (\ref{eq:psc}) can be expressed as:
\begin{equation}
  {\bf p}_{i}={\bf v}_{i}
- V_{-i}({V}_{-i}^{\prime}{V}_{-i}
-[{e}^{\prime}][{V}_{-i-i}])^{-1}
 ({V}_{-i}^{\prime}{\bf v}_{i}-{{V^{\prime}}_{-ii}}{e})
 \label{pvec}
 \end{equation}
where ${\bf p}_{i}$ is $n \times 1$ vector containing  profile-based sensitivity coefficients for $\theta_{i}$  evaluated at the $n$ prediction points, ${\bf v}_{i}$ is the corresponding vector of local sensitivity coefficients.
If the linear approximation to the model function is adequate, i.e., Hessian terms can be set to zero or when the model fits data exactly ($e=0$), the vector of profile-based sensitivity coefficients in equation (\ref{pvec}) reduces to the following:
 \begin{equation}
  {\bf p}_{i}={\bf v}_{i}
- V_{-i}({V}_{-i}^{\prime}{V}_{-i})^{-1}
 {V}_{-i}^{\prime}{\bf v}_{i}
 \label{pvec1}
 \end{equation}
which can be expressed as:
\begin{equation}
  {\bf p}_{i}=[\bf{I}_{n}-{\cal P}_{V_{-i}}]{\bf v}_{i}
 \label{project}
 \end{equation}
where $\bf{I}_{n}$ is the $n$-dimensional identity matrix and ${\cal P}_{V_{-i}}=V_{-i}({V}_{-i}^{\prime}{V}_{-i})^{-1}{V}_{-i}^{\prime}$ is the projection matrix that orthogonally projects the columns of $V_{-i}$ onto themselves. The projection matrix $\bf{I}_{n}-{\cal P}_{V_{-i}}$ projects the columns of $V_{-i}$ onto the orthogonal complement of the space spanned by the columns of $V_{-i}$.
A close examination of the formulation given in equation (\ref{project}) reveals that  ${\bf p}_{i}$ is the vector of least squares residuals obtained from linearly regressing  ${\bf v}_{i}$ on regressor variables given by $V_{-i}$. Note that the vector ${\bf v}_{i}$ can be orthogonally decomposed as:
 \begin{equation}
  {\bf v}_{i}={\cal P}_{V_{-i}}{\bf v}_{i}+[\bf{I}_{n}-{\cal P}_{V_{-i}}]{\bf v}_{i}
 \label{project1}
 \end{equation}
 If the local effect of  $\theta_{i}$ on the predicted response is highly correlated with the effects of the remaining parameters, $\Theta_{-i}$, the first term in equation (\ref{project1}) will have large magnitude compared to the second term indicating small magnitude of profile-based sensitivities. On the other hand, weak correlations among the effects of parameters on the predicted response indicate large magnitude of profile-based sensitivities. This is to say that $\bf{p}_{i}$ measures the influence that  $\theta_{i}$ exerts on the predicted response after the removal of its co-dependencies with the remaining parameters. With this understanding, profile-based sensitivities represent a particular orthogonalization of parameter space  so that the individual impacts of the resulting parameters are more independent from each other than those of original parameters $\Theta$. Inclusion of the second order derivatives in Equation (\ref{pvec}) suggests that model nonlinearity is accounted for in this particular orthogonalization.

Sulieman {\em et. al}.\cite{sulieman4} described  profile-based sensitivity procedure using the notion of model re-parametrization. They showed that the slopes of the profile traces in the re-parametrized model represent the foundation for the underlying definition of the profile-based sensitivity measure.
Sulieman {\em et al}.\cite{sulieman2} extended the profile-based sensitivity assessment to the parameter estimation in multi-response regression models. Sulieman {\em et al}.\cite{sulieman3} presented a comparative analysis of the profile-based sensitivity and Fourier Amplitude Sensitivity Test (FAST). They showed that while FAST accounts for model nonlinearity to all orders it fails to account for parameter co-dependencies which are considered in the profile-based sensitivity measure.
\section{PROFILE-BASED D-OPTIMAL DESIGN}
D-optimal designs are one of the most commonly used alphabet designs. A D-optimal design minimizes the volume of the parameter joint inference region or equivalently maximizing the determinant of the Fisher Information matrix with respect to the design settings. Box and Lucas \cite{BL59} gave the first formulation and geometric interpretation of the D-optimal design for nonlinear models. They defined the D-optimality objective function, using the local  sensitivity coefficients, as:
\begin{equation}
max D=|V_{0}^{\prime}V_{0}|
\label{Doptim}
\end{equation}
with respect to design settings, ${\bf x}$,  where the matrix of local sensitivity coefficients $V_{0}$ is evaluated at an initial parameter estimates $\Theta_{0}$.
Under the linear approximation, the model response surface is replaced by its tangent plane and the usual ellipsoidal joint inference region for $\Theta$ is the image in parameter space of a spherical region on the tangent plane. The volume of the ellipsoidal region evaluated at $\Theta_{0}$ is given by $|V_{0}^{\prime}V_{0}|^{-1/2}$.
 By maximizing $D$, this volume is minimized. Thus, for a given nonlinear model response, the D-optimal criterion ensures that the design is such that large regions on the tangent plane map into small regions in the parameter space.When model nonlinearity is pronounced, the {\it local} D-optimality can produce designs with poor performance and little information about parameters.
Hamilton and Watts \cite{HW85} introduced quadratic designs based on  second-order approximation to the volume of the inference region of $\Theta$. Quadratic designs have the distinct advantage of taking into account the nonlinearity of response function. Benabbas {\em et al}.\cite{BAM05} proposed a curvature-based method for optimal experimental design for parameter estimation in multi-response nonlinear dynamic models. Vila and Gauchi\cite{VG10} constructed non-sequential optimal designs based on the expected volume of exact parameter confidence regions. These designs generally result in repeated experiments on $k$-support points for models with $k$ parameters and tend to reduce parameter nonlinearities.
Gao and Zhou\cite{GZ17} developed a nonlinear D-optimal design criterion based on the second-order least squares estimator for regression models with asymmetric error distribution.

 Using the profile-based sensitivity coefficients defined in equation (\ref{pvec}), the profile-based  D-optimality can be defined as maximizing:
 \begin{equation}
max D_{P}=|P_{0}^{\prime}P_{0}|
\label{Dp}
\end{equation}
with respect to ${\bf x}$, where the matrix $P=\left[{\bf p}_{1} {\bf p}_{2} \ldots {\bf p}_{k}\right] $ is evaluated at $\Theta_{0}$, i.e., each element ${\bf p}_{i}$ is evaluated at $\Theta_{0}$.\\
As discussed in the previous section, profile-based sensitivity coefficients orthogonalize the parameter space so that the resulting parameters are less correlated than the original parameters. By maximizing $D_{p}$, the volume of the inference region in the less-correlated parameter space is minimized. Hence, the resulting design produce more precise and less correlated parameter estimates than the corresponding local $D$-optimal design.

The $(i,j)^{th}$ element of the $k\times k$ matrix $P^{\prime}P$ is given by:
\begin{equation}
  {\bf p}_{i}^{\prime}{\bf p}_{j}={\bf v}_{i}^{\prime}{\bf v}_{j}-{\bf v}_{i}^{'}V_{-j}H^{-1}_{-j-j}{\bf h}_{-jj}-{\bf h}^{'}_{-ii}H^{-1}_{-i-i}V^{'}_{-i}{\bf v}_{j}+{\bf h}^{'}_{-ii}H^{-1}_{-i-i}V^{'}_{-i}V_{-j}H^{-1}_{-j-j}{\bf h}_{-jj}
  \label{Dp1}
  \end{equation}
where $H_{-i-i}= {V}_{-i}^{'}{V}_{-i}-[{e}^{'}][{V}_{-i-i}]$ is $k-1$ square matrix containing Hessian terms corresponding to the parameters $\Theta_{-i}$, ${\bf h}_{-ii}={V}_{-i}^{\prime}{\bf v}_{i}-{{V^{\prime}}_{-ii}}{e}$ is $k-1$-element vector containing  Hessian terms corresponding to $\theta_{i}$ and $\Theta_{-i}$. Similarly, the matrix $H_{-j-j}$ and vector ${\bf h}_{-jj}$ contain the corresponding Hessian terms for $\Theta_{-j}$ and $\theta_{j}$. The first term in equation (\ref{Dp1}) is equal to the $(i,j)^{th}$ element of the  matrix $V_{0}^{\prime}V_{0}$ used in the local D-optimality criterion. The remaining terms are proportionate  to pairwise products of various  co-dependency  structures among the conditional parameter estimates in  $\Theta_{-i}$ and those in $\Theta_{-j}$ in addition to the  corresponding co-dependencies with the conditioning parameters  $\theta_{i}$ and $\theta_{j}$.  Thus, the objective function in $D_{P}$-optimality is equal to the objective function in the conventional $D$-optimality corrected for the correlation structures among parameters. Equation (\ref{Dp1}) suggests that maximum $D_{P}$ is obtained by maximizing $D$ (first term) and minimizing the magnitudes of the middle two terms representing correlation structures. This suggests that $D_{P}$ maximizes the information content of the design while accounting for the correlations among parameter estimates that result from an existing design or/and model formulations. These correlations are ignored by the $D$-optimal criterion.
Similar to the $D$-optimal designs, the $D_{P}$-optimal designs are invariant under nonsingular transformation of parameters.

The ability of the $D$-optimal design to estimate model parameters relative to the $D_{P}$-optimal design  is measured by its $D$-efficiency. The $D$-efficiency  is defined by:
\begin{equation}
D_{eff}=(\frac{|V_{0}^{\prime}V_{0}|}{|P_{0}^{\prime}P_{0}|})^{\frac{1}{k}}\times 100\%
\label{eff}
\end{equation}
 The $D_{eff}$ gives the percentage of the experimental effort of the $D$-optimal  design required by $D_{p}$-optimal design in order to produce parameter estimates of the same precision.
\section{Model Examples and Simulations}
To carry out the computations for constructing the optimal $D$ and $D_{p}$ designs, we initially evaluate the two criteria at a selected  set of fine grid points in the design region using initial parameter values, $\Theta_{0}$.  Then we performed a minimization algorithm  for each of $D^{-1}$ and $D_{p}^{-1}$ using one of the two multivariable nonlinear minimization routines found in MATLAB's Optimization Toolbox: fminsearch for unconstrained optimization and fmincon for constrained optimization. The starting  point ${\bf x}_{0}$ used for the minimization is the point at which $D$ and $D_{p}$ are maximized in the initial grid search. To ensure that the resulting optimal design point is global one, a further grid search exploring the interior of the design region is carried out. A simulation study is conducted for each model example in order to evaluate the performance of the constructed designs.
\subsection{Michaelis-Menten Enzyme Kinetic Model}
The Michaelis-Menten model is one of the most widely used models in the biological sciences. It is commonly used in enzymatic kinetics with
well-known  formulation:
\begin{equation}
{\bf y} =\frac{\theta_{1}x}{\theta_{2} +x}+ \mbox{\boldmath $ \epsilon $}
\label{menten}
\end{equation}
where $y$ is the measured initial velocity  of an enzymatic
reaction and $x$ is the substrate concentration. The unknown
parameters $\theta_{1}$ and $\theta_{2}$ represent maximum
conversion rate and Michaelis-Menten constant, respectively.\\
One of the most widely data sets used to fit the model is published in Bates and Watts\cite{BW88} representing reaction velocity measurements  with enzyme treated with Puromycin and with untreated enzyme. Table 1 depicts the design and velocity values for the treated enzyme.
\begin{table}[ht]
\centering
{\renewcommand{\arraystretch}{1}
Table 1: Data set for the Michaelis-Menten model reported in Bates \& Watts \cite{BW88}\\
\vspace{.12in}
\begin{tabular}{ccccccc} \hline\hline
& & {\small Observation  no.} & {\small Substrate Concentration}{\tiny (ppm)}  & {\small Velocity (Treated)}{\tiny (counts/min$^{2}$)} & & \\ \hline
& & 1 & 0.02 & 76 & & \\
& & 2 & &     47 & & \\
& & 3 & 0.06 & 97 & & \\
& & 4 & & 107 & & \\
& & 5 & 0.11 & 123 & & \\
& & 6 & & 139 & & \\
& &  7 & 0.22 & 159 & & \\
& & 8 & & 152 & & \\
& & 9 & 0.56 & 191 & & \\
& & 10 & & 201 & & \\
& & 11 & 1.10 & 207 & & \\
& & 12 & & 200 & & \\ \hline
\end{tabular}}
\label{table:past.dat}
\end{table}
\begin{figure}[t]
\setlength{\unitlength}{0.6in} \centerline{
\begin{picture}(6.6,6.6)(0.9,-0.7)
%\linethickness{.5pt}
\epsfysize=6in
\includegraphics[width=.80 \textwidth]{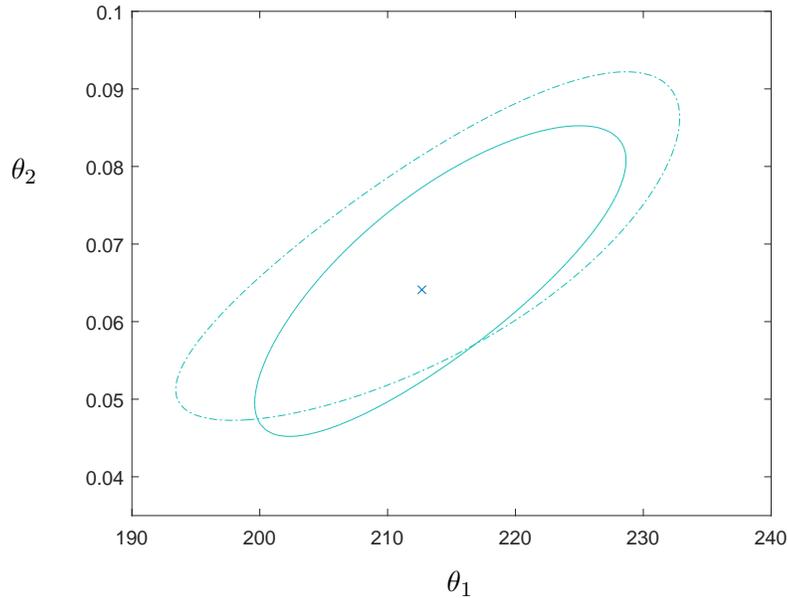}
\put(-7.2,3.6){\makebox(0,0){{\small ${\theta}_{2}$}}}
 \put(-3.4,0){\makebox(0,0){{\small ${\theta}_{1}$}}}
\end{picture}}
\vspace{-0.35in}
 {\renewcommand{\baselinestretch}{1.0}
\caption{$90\% $ joint likelihood regions for the parameter in the Michaelis-Menten model based on unconditional least squares estimation (solid line) and conditional least squares estimations (dashed-dotted line). The unconditional least squares estimate ($\hat{\theta}_{1}=212.68$, $\hat{\theta}_{2}=0.064$) is indicated by x.}}
\end{figure}
This original design consists of six different experimental settings, each with two replications. We used the data to  construct the joint inference regions for the parameter estimates of $\theta_{1}$ and $\theta_{2}$, $ \hat{\theta}_{1}$ and $\hat{\theta}_{2}$. Figure 1 depicts the 90\% likelihood inference regions for the two parameters based on unconditional and conditional least squares estimation of the model. The unconditional likelihood contour (solid line) was determined by evaluating the  sum of squares function for an arbitrary grid of ($\hat{\theta}_{1}$, $\hat{\theta}_{2}$) values in their respective ranges of uncertainty. Whereas, for the conditional likelihood contour (dashed-dotted line)  the sum of squares function was evaluated using a grid points ($\tilde{\theta}_{1}(\theta_{2})$, $\tilde{\theta}_{2}(\theta_{1})$) where $\tilde{\theta}_{1}(\theta_{2})$ is the estimate of $\theta_{1}$ conditional on selected values of ${\theta}_{2}$ in its range of variation and $\tilde{\theta}_{2}(\theta_{1})$ is the least squares estimate of $\theta_{2}$ conditional on selected values of ${\theta}_{1}$ in its range of variation.\\
It is evident from Figure 1 that the 90\% conditional likelihood inference region is less elliptical with reduced inclination reflecting decreased correlation between the two parameters. The conditional estimation procedure reduces the correlations induced by the simultaneous search for optimal least squares estimates in  $\theta_{1}$ and $\theta_{2}$ directions.   Hence, the $D_{p}$-optimality provides a basis for designing experimental settings that produce less correlated parameter estimates.
\subsubsection{Starting Design}
Using the local D-optimal criterion in equation (\ref{Doptim}), Bates and Watts\cite{BW88} constructed a starting design for the model in equation (\ref{menten}). They showed that the design does not depend on the conditionally linear parameter $\theta_{1}$ and used $\theta_{2}^{0}=0.1$ as  starting value of $\theta_{2}$. The maximum $D$ occurred at $x_{1}=1.1$ and $x_{2}=0.085$ where the value 1.1 is the maximum concentration reported in the original design given in Table 1  and 0.085 is nearly the half-concentration in the same design. They compared their design to the original design and concluded that their design produced smaller linear approximation inference region for $\theta_{1}$ and $\theta_{2}$ with lower correlation between the parameter estimates.
\begin{figure}[t]
\setlength{\unitlength}{0.6in} \centerline{
\begin{picture}(6.6,6.6)(0.9,-0.7)
%\linethickness{.5pt}
\epsfysize=6in
\includegraphics[width=.80 \textwidth]{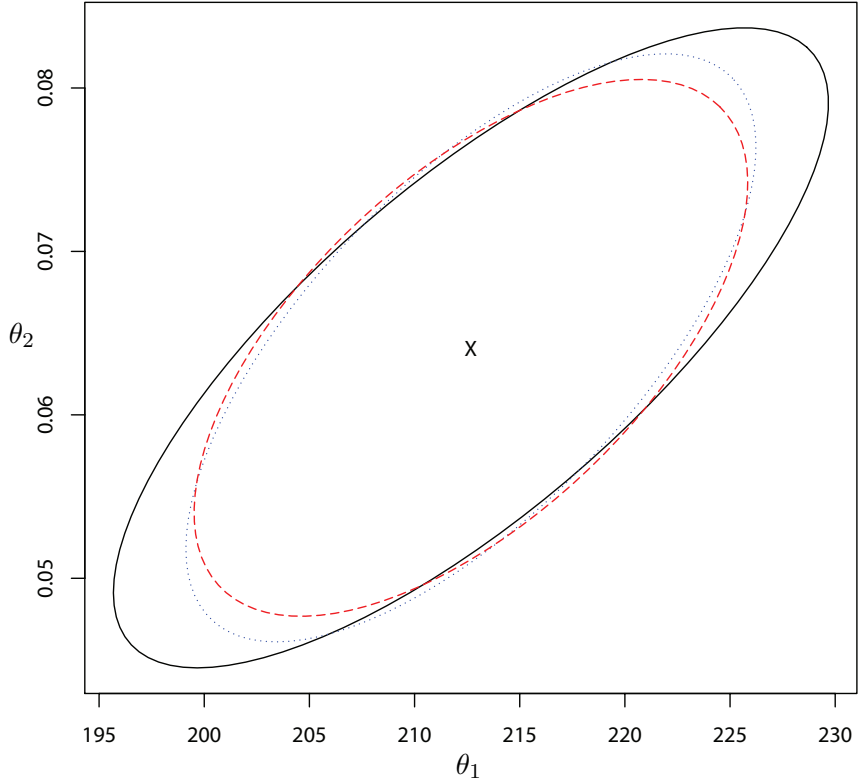}
\put(-7.3,3.6){\makebox(0,0){{\small ${\theta}_{2}$}}}
 \put(-3.4,-.16){\makebox(0,0){{\small ${\theta}_{1}$}}}
\end{picture}}
\vspace{-0.35in}
 {\renewcommand{\baselinestretch}{1.0}
\caption{$95\% $ approximate confidence region for three designs for estimating the Michaelis-Menten model parameters ${\theta}_{1}$ and ${\theta}_{2}$. The largest region (solid line) represents the original design (Table 1), the middle region (dotted line) represents the $D$- optimal design and the smallest region (dashed line) represents the $D_{p}$-optimal design }}
\end{figure}

In constructing the starting $D_{P}$ design, we set the unknown residual vector in equation (\ref{pvec}) to zero and so the ${\bf p}_{i}$ vectors are reduced to the form given in equation (\ref{project}). We write the model equation (\ref{menten}) as $f(x,\Theta) =\theta_{1}g(x,\theta_{2})$ where $g(x,\theta_{2})=\displaystyle  \frac{x}{\theta_{2} +x}$. For notational convenience, we write $g(x,\theta_{2})=g$ and so $f(x,\Theta) =\theta_{1}{\bf g}$. It can be shown that the $2$-element vectors of profile-based sensitivity of $\theta_{1}$ and $\theta_{2}$ are given by:
\begin{equation}
{\bf p}_{1}=[I_{2}-{\bf g}_{\theta_{2}}({\bf g}^{\prime}_{\theta_{2}}{\bf g}_{\theta_{2}})^{-1}{\bf g}^{\prime}_{\theta_{2}}]{\bf g}_{\theta_{2}}
\end{equation}
\begin{equation}
{\bf p}_{2}=\theta_{1}[I_{2}-{\bf g}({\bf g}^{\prime}{\bf g})^{-1}{\bf g}^{\prime}]{\bf g}\propto [I_{2}-{\bf g}({\bf g}^{\prime}{\bf g})^{-1}{\bf g}^{\prime}]{\bf g}
\end{equation}
where ${\bf g}=(g(x_{1},\theta_{2}), g(x_{2},\theta_{2}))^{\prime}$ and ${\bf g}_{\theta_{2}}=\displaystyle (\frac{\partial{g(x_{1},\theta_{2})}}{\partial \theta_{2}}, \frac{\partial{g(x_{2},\theta_{2})}}{\partial \theta_{2}})^{\prime}$.
It is obvious that ${\bf p}_{1}$ is free of the conditionally linear parameter $\theta_{1}$ and that $\theta_{1}$ appears linearly in ${\bf p}_{2}$ as a proportion constant to a term that depends on $\theta_{2}$ only. This implies that the optimum value of $D_{P}$ is independent of $\theta_{1}$. As shown by Bates and Watts\cite{BW88},the classical $D$-optimal designs are independent of the conditionally linear parameters for most nonlinear models.  This property  holds also true for  the $D_{P}$-optimal designs only when models are intrinsically linear (${V}_{-i-i}$ and ${V}_{-ii}$ can be set to zero) or when data fits the model perfectly ($e$ is zero).

Using MATLAB 8.2.0 optimizer, the locations of maximum $D_{P}$ were found at $x_{1}=1.1$ and $x_{2}=0.056$. In comparing the resulting  $D_{P}$-optimal design to the $D$-optimal design constructed by Bates and Watts\cite{BW88}, we places six replications at $x_{1}=1.1$ and six replications at $x_{2}=0.056$. The corresponding $D$-optimal design consists of six replications of  $x_{1}=1.1$ and six replications of $x_{2}=0.085$. As shown in Table 1, the original design consists of six design points, each has two replications. Figure 2 shows the $95\%$ linear approximation confidence regions for $\theta_{1}$ and $\theta_{2}$ using the three designs assuming that all designs produced same parameter estimates and residual variance.

 We can clearly see that the $D_{P}$-optimal design  produces the smallest joint confidence region (dashed line) and smaller confidence intervals. Also the correlation between the two parameter estimates is the lowest (0.65) for the $D_{P}$-optimal design compared to 0.68 for the $D$-optimal design and 0.76 for the original design. The $D$-efficiency is $95\%$ indicating that $95\%$ of the optimal $D$ experimental effort is needed by the optimal $D_{P}$ design in order to produce similar accuracy of parameter estimates. In other words, the $D$ optimal design requires $5\%$ more of experimental effort in order to obtain as accurate parameter estimates as that given by the $D_{P}$ design.
 \subsubsection{Sequential Design}
  When some experiments are already done, a natural way of designing an experiment is to use a sequential method. Sequential designs are appealing because they offer the chance to change strategy after the first round of experiments has been completed and new information is available. Of particular interest is the case when one additional design point is desired. In $D$-optimality, this is achieved by maximizing $|V_{n+1}^{\prime}V_{n+1}|$ with respect to the $(n+1)^{th}$ design point, $\bf{x}_{n+1}$, where
  \begin{equation}
%\begin{eqnarray*}
V_{n+1}=\left[ \begin{array}{c}
V_{n} \\ {\bf v}_{n+1}
\label{sequential}
\end{array}   \right]
%\end{eqnarray*}
\end{equation}
where $V_{n}$ is the design matrix consisting of the local sensitivity measures for the pre-existing experimental settings and ${\bf v}_{n+1}$ is $k$-element vector  of sensitivity coefficients that correspond to the new experimental setting being selected.
The corresponding profile-based sequential design strategy maximizes $|P_{n+1}^{\prime}P_{n+1}|$ where
\begin{equation}
P_{n+1}=\left[ \begin{array}{c}
P_{n} \\ {\bf p}_{n+1}
\label{sequential}
\end{array}   \right]
\end{equation}
 and ${\bf p}_{n+1}$ is $k$-element vector  of profile-based sensitivity coefficients corresponding to the $(n+1)^{th}$ experimental setting. $V_{n+1}$ and $P_{n+1}$ are evaluated at the least squares parameter estimates from the already existing $n$ experiments.

The original design for the Michaelis-Menten model given in Table 1 was used to obtain the least squares estimates of $\theta_{1}$ and $\theta_{2}$. These estimates were used to evaluate $V_{n+1}$ and $P_{n+1}$ above.
The $13^{th}$ concentration point is generated using MATLAB 8.2.0 optimizer for restricted $x$, $0<x \leq x_{max}=1.1$. The optimal value for the additional concentration point is $x=0.0747$ when $D$ is maximized and $x=0.05116$ when $D_{P}$ is maximized. With one additional design point, the efficiency for the new $D_{P}$ design is $98\%$.
\newpage
\noindent {\bf Simulation Study}\\

In a simple attempt to evaluate the information content of the 13-point design, formed by adding the new optimal value of $x$ to the existing design in Table 1, the parameters  $\theta_{1}$ and $\theta_{2}$ are re-estimated twice: one time using the new design derived by $D$ criterion and the other using the design derived by $D_{P}$ criterion. The response variable for the $13^{th}$ optimal concentration point is simulated using the fitted model obtained from the original 12-point design and adding normally
distributed random noise. The estimation procedure for each design was carried out for 2000 simulations. For each simulation, the estimated linear-approximation based  variance-covariance matrix, $s^{2}(V^{\prime}V)^{-1}$ is evaluated at the least squares estimate $\hat{\Theta}$ and recorded.
\begin{figure}[t]
\setlength{\unitlength}{0.7in} \centerline{
\begin{picture}(6,6)(0,-1.8)
%\linethickness{.5pt}
\epsfysize=6in
\includegraphics[width=.80 \textwidth]{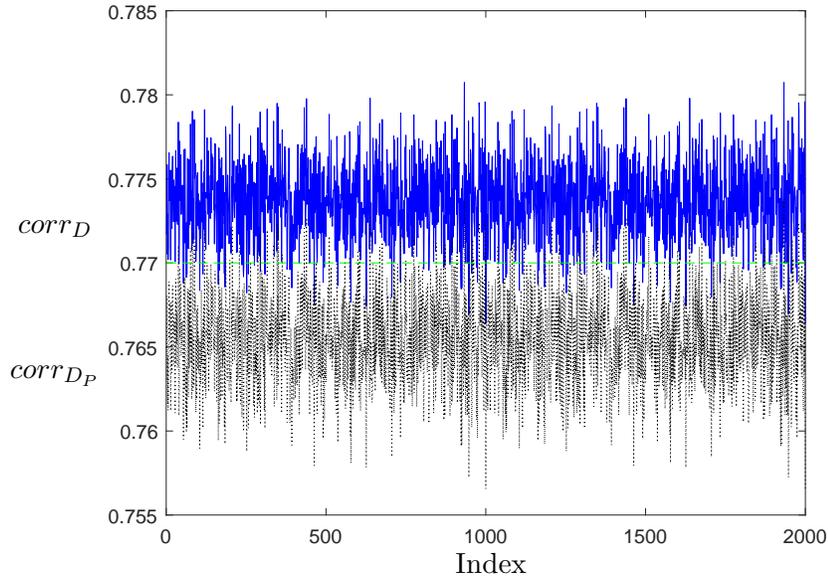}
\put(-2.94,0.15){\makebox(0,0){{\small Index}}}
\put(-6.2,2.65){\makebox(0,0){{\small $corr_{D}$}}}
 \put(-6.2,1.55){\makebox(0,0){{\small $corr_{D_{P}}$}}}
\end{picture}}
\vspace{-1.3in}
 {\renewcommand{\baselinestretch}{1.0}
\caption{Simulated correlation coefficients between $\hat{\theta}_{1}$ and $\hat{\theta}_{2}$ in Michaelis-Menten Model using 13-points sequential design resulted from $D$-optimality (solid line) and  $D_{P}$-optimality (dotted line). The dashed line gives the correlation coefficient (0.77) from the 12-point original design}}
\label{corr}
\end{figure}
\begin{figure}[t]
\setlength{\unitlength}{1.0in} \centerline{
\begin{picture}(4,4)(0.5,-1.5)
%\linethickness{.5pt}
\epsfysize=6in
\includegraphics[width=.70 \textwidth]{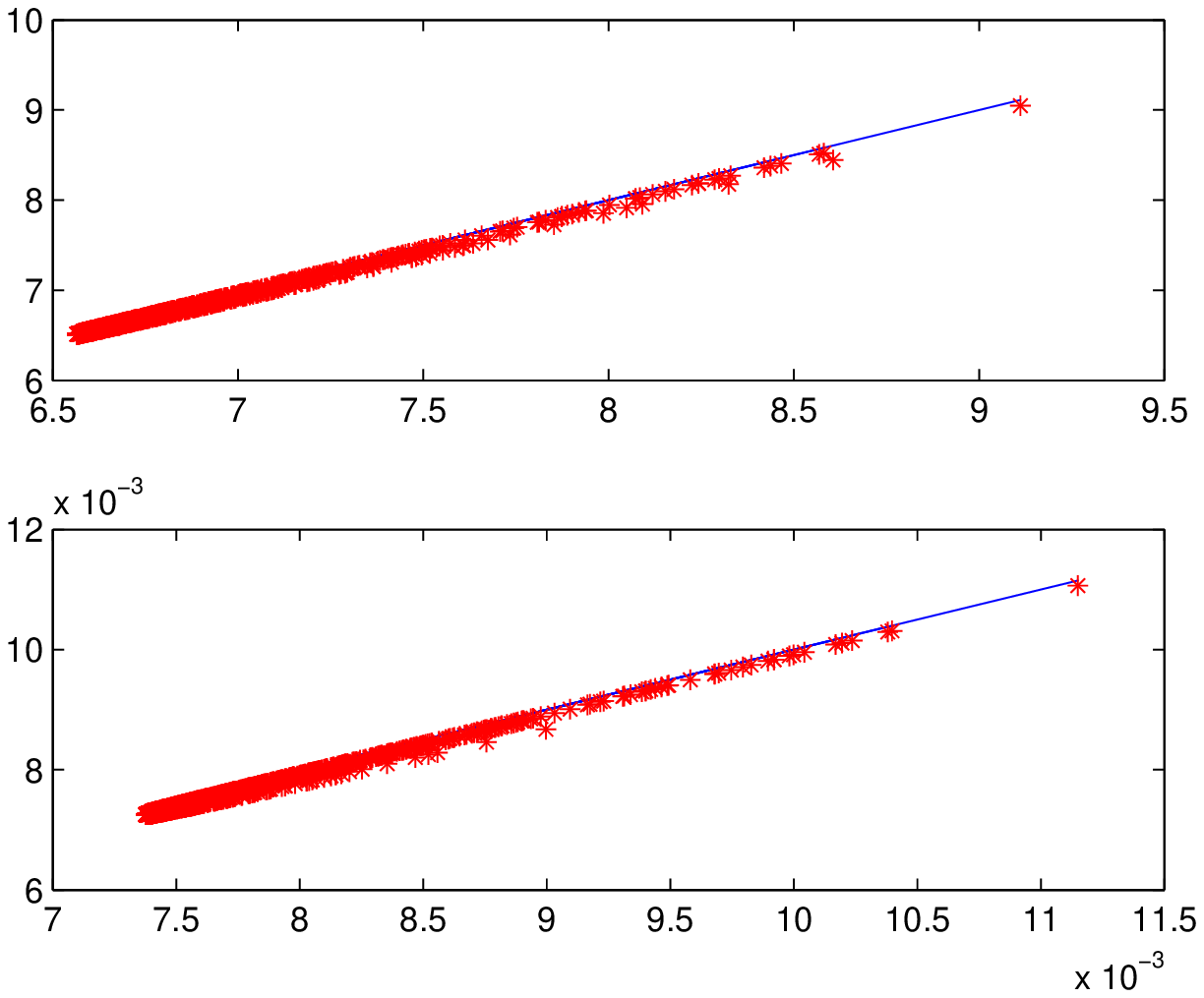}
\put(-1.9,0.05){\makebox(0,0){{$se_{D}(\hat{\theta}_{2})$}}}
\put(-2.0,1.45){\makebox(0,0){{ $se_{D}(\hat{\theta}_{1})$}}}
\put(-3.7,2.0){\makebox(0,0){{ $se_{D_{P}}(\hat{\theta}_{1})$}}}
 \put(-3.7,0.8){\makebox(0,0){{$se_{D_{P}}(\hat{\theta}_{2})$}}}
\end{picture}}
\vspace{-1.45in}
 {\renewcommand{\baselinestretch}{1.0}
\caption{Simulated standard errors of $\hat{\theta}_{1}$ and $\hat{\theta}_{2}$ in Michaelis-Menten model using the 13-point sequential design resulted from $D$-optimality and  $D_{P}$-optimality (asterisk points). The solid line is the line of equality}}
\end{figure}

Figure 3 shows  the resulting correlation coefficients between $\hat{\theta}_{1}$ and $\hat{\theta}_{2}$ plotted against the simulation number. The dotted line represents the correlation coefficient resulting from the $D_{P}$ design while the solid line represents the corresponding correlation resulting from the $D$ design. The horizontal dashed line gives the correlation coefficient (0.77)  resulted from the original design given in Table 1. The line charts of Figure 3 clearly demonstrate that, for vast majority of the simulations,  the $D_{P}$ design gives reduced correlation coefficient from that given by the original design and these correlations are remarkably lower than correlations generated from the $D$ design.\\
Furthermore, in Figure 4 we present scatter plots of the estimated standard errors of $\hat{\theta}_{1}$ and $\hat{\theta}_{2}$ from the 2000 simulations. The parameter standard errors resulted from $D$ design ($se_{D}$) are plotted on the horizontal axis while the $D_{P}$ design standard errors ($se_{D_{p}}$) are plotted (asterisk points) on the vertical axis. The solid line is the line of equality of $se_{D}$ and $se_{D_{P}}$ so that it is easily seen that in a significant portion of the simulations the $se_{D_{P}}$ is less than  $se_{D}$, reflecting higher accuracy of parameter estimates. The average value of the simulated D-efficiency scores is  $97.6\%$ which suggests that the two designs are of comparable efficiency in estimating the two parameters. The $D_{P}$ design, however, has the distinctive ability to increase parameter estimate precision while reducing their correlations.
\subsection{Hougen-Watson Model}
The Hougen-Watson model is common in chemical kinetics of catalyzed reactions. It expresses the reaction rate in terms of the catalyst variables, the temperature and concentrations of reactants. One of the expressions the model takes is:
\begin{equation}
 f({\bf x},\Theta) =\frac{\theta_{1}\theta_{3}(x_{2}-x_{3}/1.632)}{1+\theta_{2}x_{1}+\theta_{3}x_{2}+\theta_{4}x_{3}}
\label{isom}
\end{equation}
where $x_{1}$, $x_{2}$ and $x_{3}$ represent the partial pressures of the reactants.
The data set, reported in Bates and Watts\cite{BW88}, consists of 24 runs of the design variables $(x_{1}, x_{2}, x_{3})$. The model was fitted to this initial 24-point design. Table 2 below shows the results.
\begin{table}[ht]
\centering
{\renewcommand{\arraystretch}{1}
Table 2: Summary of parameter estimates  for the
Hougen-Watson model \vspace{.2in}
\begin{tabular}{ccccccccccc} \hline\hline
& & Parameter & Estimate & St.error & &  Correlation  & & & & \\ \hline
& & $\theta_{1}$ & 35.92 & 8.21 & & 1.00  &  & & & \\
& & $\theta_{2}$ & 0.071 & 0.178 & & -0.805 & 1.00 & & & \\
& & $\theta_{3}$ & 0.038 & 0.099 & & -0.840 & 0.998 & 1.00 & & \\
& & $\theta_{4}$ & 0.167 & 0.415 & & -0.790 & 0.998 & 0.995 & 1.00 & \\ \hline
\end{tabular}}
\end{table}

The correlations between the parameter estimates are clearly very high and the model suffers from pronounced nonlinearity, Bates and Watts\cite{BW88}.\\
 To generate an additional design point to improve parameter precision, Bates and Watts\cite{BW88} used the $D$-optimality with design variables restricted to the following ranges: $100\leq x_{1}\leq 400$, $75 \leq x_{2} \leq 350$ and $30 \leq x_{3} \leq150$. They searched for the maximum $D$ at the corner points of the restricted design region and  inside its interior. They found the combination $(x_{1}=100, x_{2}=350, x_{3}=30)$ to be the point at which the optimum $D$ occurred and hence they recommended this corner point for the next experimental run.

We implemented  the same strategy using MATLAB 8.2.0 optimizer. At first, we  evaluated the $D_{P}$ criterion using equation (\ref{pvec}) at the corner points of the design region above and at the original 24 design points. The maximum $D_{P}$ occurred at the design point $(x_{1}=251, x_{2}=294, x_{3}=41.5)$. We used this point as starting point for the maximization of $D_{P}$. The $25^{th}$ design point at which the optimum value was found is $(x_{1}=245, x_{2}=300, x_{3}=40)$.\\
\vspace{.1in}

\noindent {\bf Simulation Study}\\

Similar to the previous example, we ran 2000 simulations of the Hougen-Watson model estimation in order to evaluate the information content of each of the two 25-point designs constructed by $D$ and $D_{P}$ criteria. In each simulation, the response variable for the additional point; $(x_{1}=100, x_{2}=350, x_{3}=30)$ for $D$ and $(x_{1}=245, x_{2}=300, x_{3}=40)$ for $D_{P}$; was estimated by using the fitted model in Table 2 and adding randomly generated noise from a normal distribution. The results are depicted in Figures 4 and 5.
\begin{figure}[t]
\setlength{\unitlength}{0.7in} \centerline{
\begin{picture}(6,6)(0,-1.8)
%\linethickness{.5pt}
\epsfysize=6in
\includegraphics[width=.80 \textwidth]{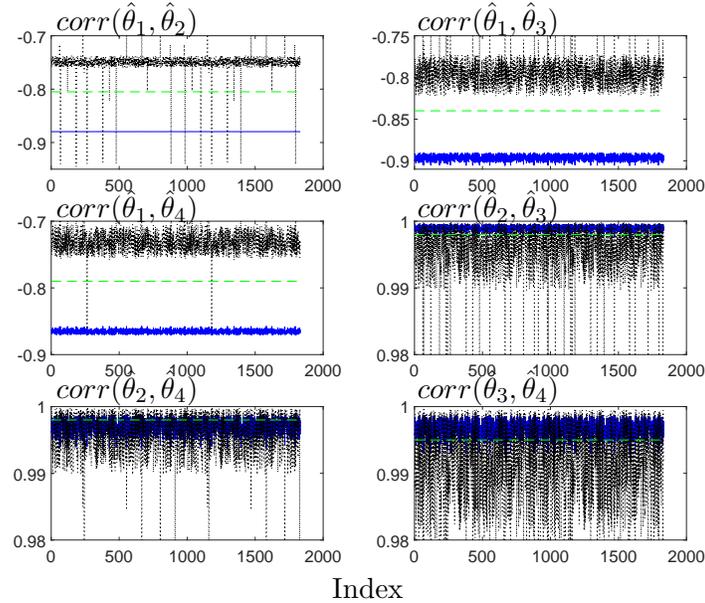}
\put(-3,0.15){\makebox(0,0){{\small Index}}}
\put(-4.8,4.4){\makebox(0,0){{\small $corr(\hat{\theta}_{1},\hat{\theta}_{2})$}}}
 \put(-2.1,4.4){\makebox(0,0){{\small $corr(\hat{\theta}_{1},\hat{\theta}_{3})$}}}
\put(-4.8,3.0){\makebox(0,0){{\small $corr(\hat{\theta}_{1},\hat{\theta}_{4})$}}}
\put(-2.1,3.0){\makebox(0,0){{\small $corr(\hat{\theta}_{2},\hat{\theta}_{3})$}}}
\put(-4.8,1.65){\makebox(0,0){{\small $corr(\hat{\theta}_{2},\hat{\theta}_{4})$}}}
\put(-2.1,1.65){\makebox(0,0){{\small $corr(\hat{\theta}_{3},\hat{\theta}_{4})$}}}
\end{picture}}
\vspace{-1.4in}
 {\renewcommand{\baselinestretch}{1.0}
\caption{Simulated correlation coefficients between the four parameter estimates in the Hougen-Watson model. The solid line represent correlations from the $D$-optimal design, the dotted line represent correlations from the $D_{P}$-optimal design and the horizontal dashed line gives the correlation coefficient from the 24-point original design}}
\label{isom1}
\end{figure}
\begin{figure}[t]
\setlength{\unitlength}{0.7in} \centerline{
\begin{picture}(6,6)(0,-1.8)
%\linethickness{.5pt}
\epsfysize=6in
\includegraphics[width=.80 \textwidth]{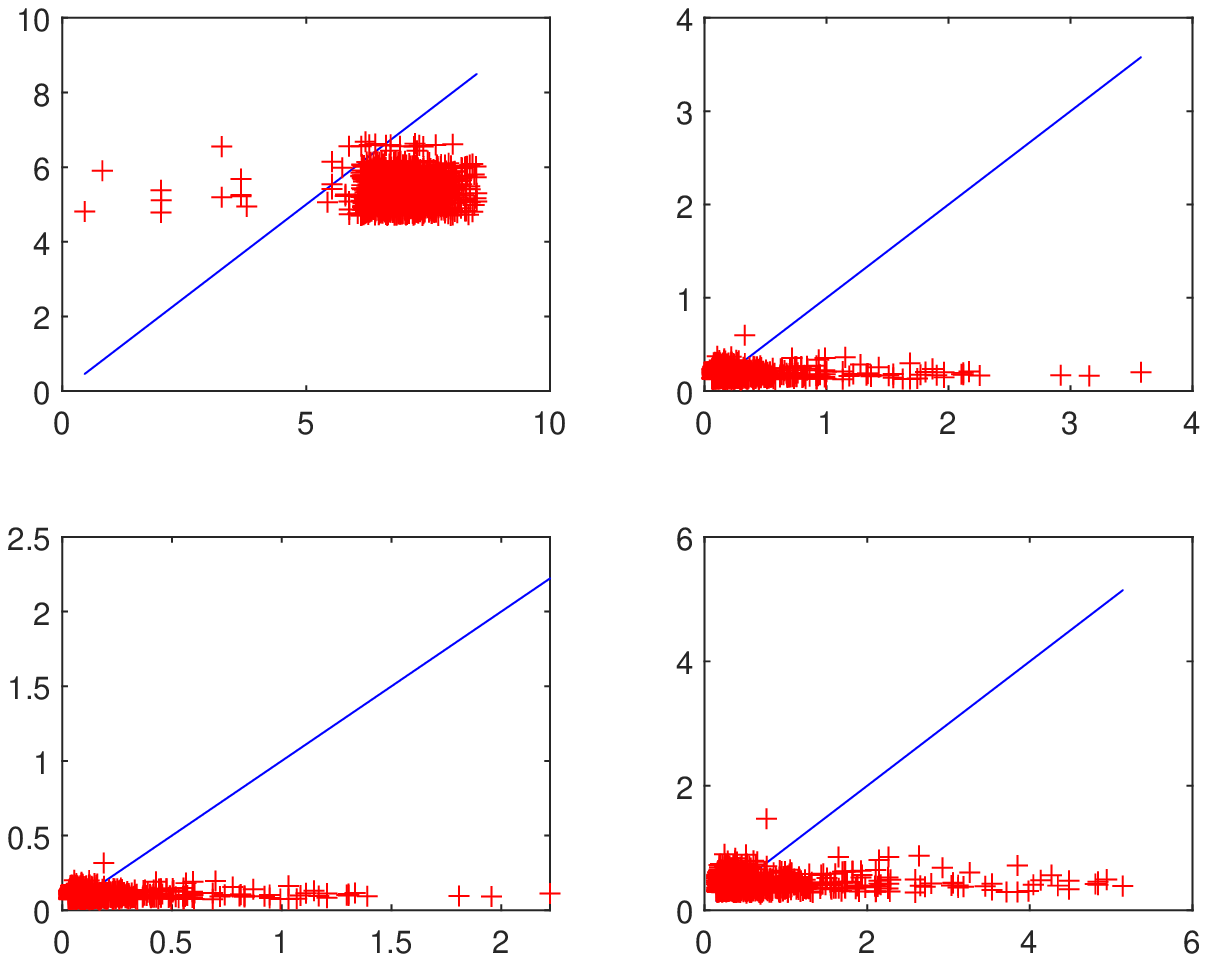}
\put(-4.4,2.4){\makebox(0,0){{\small $\hat{se}_{1_{D}}$}}}
\put(-5.9,3.5){\makebox(0,0){{\small $\hat{se}_{1_{D_{P}}}$}}}
 \put(-1.5,2.4){\makebox(0,0){{\small $\hat{se}_{2_{D}}$}}}
\put(-3.0,3.5){\makebox(0,0){{\small $\hat{se}_{2_{D_{P}}}$}}}
\put(-4.2,0.3){\makebox(0,0){{\small $\hat{se}_{3_{D}}$}}}
\put(-5.9,1.3){\makebox(0,0){{\small $\hat{se}_{3_{D_{P}}}$}}}
 \put(-1.5,0.3){\makebox(0,0){{\small $\hat{se}_{4_{D}}$}}}
\put(-3.0,1.3){\makebox(0,0){{\small $\hat{se}_{4_{D_{P}}}$}}}
\end{picture}}
\vspace{-1.4in}
 {\renewcommand{\baselinestretch}{1.0}
\caption{Simulated standard errors of $\hat{\theta}_{1}$, $\hat{\theta}_{2}$, $\hat{\theta}_{3}$ and $\hat{\theta}_{4}$  in the Hougen-Watson model using the 25-points sequential design resulted from $D$-optimality and  $D_{P}$-optimality (asterisk points). The solid line is the line of equality}}
\label{isom2}
\end{figure}

Figure 4 clearly demonstrates that, for vast majority of the simulations, the magnitudes of the correlations among parameter estimates are lower for the $D_{P}$ (dotted line) than the $D$ design (solid line). The reduction in correlations is most pronounced for relationships involving $\hat{\theta}_{1}$. This is to say that the location of the additional design point generated by the $D_{P}$ criterion provides informative experimental setting to reduce dependencies between $\theta_{1}$ from the other parameter estimates, thereby, reducing the overall ill-conditioning of the model estimation.  The additional design point generated by the classical $D$ criterion gave rise to higher  magnitudes of all correlations involving $\hat{\theta}_{1}$. As for the correlations involving $\hat{\theta}_{2}$, $\hat{\theta}_{3}$ and $\hat{\theta}_{4}$, the $D_{P}$ continued to produce lower values than those produced by the $D$ design. Because the $D_{P}$ criterion accounts for the correlation structure among parameters based on second-order derivative information, the resulting correlations from  $D_{P}$ design are seen to have more volatility that the correlations  produced by the classical $D$ design.

Figure 5 shows the scatter plots of the estimated standard errors of of the parameter estimates. The solid straight is the line of equality of the the estimated standard errors produced by the $D_{P}$ design,$se_{D_{P}}$, and the corresponding standard errors produced by the $D$ design, $se_{D}$. Clearly, the simulated standard errors of the four parameter estimates are lower for the  $D_{P}$ design (asterisk points) than that for the $D$ design in substantially all the simulations. It should be pointed out that, except for $\hat{\theta}_{1}$,  the simulated standard errors of the other three parameter estimates by the $D$ and $D_{P}$ designs are in general larger than those reported in Table 2 above. This is to say that the additional experimental setting in each design was not informative enough  to improve the precision of $\hat{\theta}_{2}$, $\hat{\theta}_{3}$ or $\hat{\theta}_{4}$. However, the additional design point in both $D$ and $D_{P}$ produced a significant improvement in the precision of $\hat{\theta}_{1}$ in nearly all of the simulations. The additional design point for the  $D_{P}$ optimal design was significantly informative in reducing the correlations involving $\hat{\theta}_{1}$  and improving its precision.
\section{CONCLUSIONS}
In this article we have applied the profile-based sensitivity coefficients developed by Sulieman et.al. \cite{sulieman1} in designing experiments for nonlinear models. Given that the profile-based sensitivity coefficients account for both parameter correlations and model nonlinearity, it has been shown that utilizing them in the $D$-optimal criterion generates more informative experimental settings. Two model examples have been used to demonstrate the computational aspects of the profile-based $D$-optimal criterion. Furthermore, simulation studies have shown that the constructed profile-based optimal designs are more efficient and informative than the classical $D$-optimal designs.
Future work will include establishing more detailed theoretical framework for the proposed design criterion and conducting further comparisons with existing nonlinear design criteria including that of Hamilton and Watts \cite{HW85} and  Vila and Gauchi \cite{VG10}.
\section{ACKNOWLEDGMENT}
The authors gratefully acknowledge the financial support of the
American University of Sharjah, United Arab Emirates.
%\newpage

\end{document}